\documentclass[final,5p,times,twocolumn]{elsarticle}
\usepackage[ruled,linesnumbered]{algorithm2e}
\usepackage{subfigure}
\usepackage{amssymb}
\usepackage{amsmath}
\usepackage{xurl}
\usepackage{caption}
\usepackage{graphicx}   
\usepackage{xcolor}     
\usepackage{booktabs}   
\usepackage{array} 

\journal{Journal}

\newcommand\fig{Fig.}
\newcommand\tab{Tab.}
\newcommand\alg{Algorithm.}
\newcommand\eqt{Eq.}
\newcommand\sn{LSN~}
\newcommand\Sec{\S}

\begin{document}

\begin{frontmatter}

\title{Monte Carlo Throughput Estimation in Dynamic LEO Satellite Networks}

\author[1]{Xiangtong Wang}



\ead{wxton@foxmail.com}

\begin{abstract}
	This study introduces a comprehensive framework for analyzing capacity dynamics and throughput performance in Low Earth Orbit satellite networks (LSNs) with unreliable inter-satellite links (ISLs). It addresses two critical limitations of prior models: 
	(1) the lack of capacity and throughput evaluation under dynamic network conditions, an inherent and unavoidable characteristic of LSNs, and 
	(2) the inability of flow-network models (e.g., maximum flow) to estimate throughput under specific, practical routing schemes, as they inherently assume traffic is routed according to an optimal flow distribution. 
To address these challenges, we propose the \textbf{Cap}acity model under \textbf{D}ynamic LSN (\textbf{Cap-DLSN}), which characterizes the time-varying link capacity under stochastic ISL failures, along with a \textbf{M}onte \textbf{C}arlo \textbf{T}hroughput \textbf{E}stimation (\textbf{MCTE}) framework that evaluates network throughput under dynamic traffic and diverse routing policies.
Simulations on operational constellations show that MCTE provides realistic and computationally efficient performance benchmarks, offering practical guidance for routing optimization or network design in next-generation LSNs.
\end{abstract}

\begin{keyword}
Capacity modeling \sep Monte Carlo throughput estimation \sep LEO satellite networks \sep Unreliable inter-satellite links
\end{keyword}

\end{frontmatter}

\section{Introduction}

Low Earth Orbit Satellite Networks (LSNs) consist of thousands of satellites designed to provide global internet services and have attracted considerable attention in recent years.
As a complementary infrastructure to terrestrial networks, LSNs have the potential to provide ubiquitous internet connectivity to 2.7 billion unserved users in remote areas, developing countries, and aviation and maritime environments.
Unlike traditional geosynchronous satellite communication systems\cite{chinasat,viasat} or delay-tolerant satellite networking approaches\cite{caini2011delay,el2020mars}, modern LSN architectures leverage densely deployed satellites with inter-satellite links (ISLs) to provide low-latency, full-coverage, and high-bandwidth services. 
Recent commercial space ventures have demonstrated the feasibility of LSNs \cite{starlink,oneweb,telesat,kuiper}, reshaping global connectivity paradigms.
As LSN deployments scale globally, rigorous quantification of system capacity and throughput has become indispensable for evaluating service availability and reliability.


The quantification of capacity and throughput in LSNs has been applied to many studies.
Recent works have focused on using maximum throughput or network capacity to guide specific tasks, including ISL planning \cite{liu2017capacity,lan2024inter}, topology design \cite{lin2022inter,guo2024constellation,jiang2020reinforcement,lai2024your}, ground station (GS) deployment\cite{hauri2020internet}, routing\cite{sun2004routing,liu2022maximum,yang2021maximum} and traffic engining\cite{wu2025sate,miao2024megate}.
Some of them define the throughput upper bound as the network capacity and formulate the capacity quantification as an optimization problem, calculating the network capacity by solving for the maximum value \cite{pachler2021updated,lan2024inter,basak2023exploring}. 
However, this definition leads to different network capacities under different routing schemes or traffic loads, which fails to reflect the inherent characteristics of the LSNs.
Specifically, Portillo and Pachler \cite{pachler2021updated,del2019technical} defined network throughput based on the Maximum Flow (MF)problem and provided load-throughput curves for public constellations. 
However, overlapping source and sink nodes in the traffic cause infinite flows during the calculation process. Even when the bandwidth of virtual super-links is constrained, this modeling artifact still leads to an overestimation of throughput (see \Sec\ref{exp:comp}).
Some studies \cite{wu2025sate,miao2024megate} modeled LSN throughput as a multi-commodity flow (MCF) problem, where different commodities represent different traffic demands, thereby partially isolating infinite flows or overestimation caused by source and sink overlaps. 
However, they inherently assume that the traffic flow is routed according to this maximum flow distribution, which therefore intrinsically unable to assess the actual aggregate throughput achievable under a specific routing scheme, such as Shortest Path First.
To bridge this gap, Basak et al. modified the MCF problem by limiting feasible traffic paths to a pre-computed set generated by a specific routing algorithm \cite{basak2023exploring,basak2025leocraft}. 
While this adaptation is a step towards routing-decoupled evaluation, it inherits the significant computational complexity as the network sizes grow \cite{abuzaid2021contracting} and remains a form of constrained optimization.


Collectively, most of these flow‑based models calculate the maximum theoretical throughput for a given traffic demands, this value does not reflect the practical traffic that can be carried under a specific routing scheme.Moreover, these methods typically assume a static network, whereas real‑world LSNs are inherently dynamic due to intermittent ISL availability.
This inherent instability causes network capacity to fluctuate over time, a dimension not captured by current capacity quantification approaches \cite{pachler2021updated,wu2025sate,basak2023exploring,basak2025leocraft}. 

To address this gap, we need a performance‑evaluation framework that accounts for the dynamic interaction between fluctuating network capacity and time‑varying traffic loads. 
We recognize that LSN capacity is fundamentally determined by infrastructure capability (e.g., link capacity, onboard data processing rate) rather than operational policies such as routing scheme or traffic pattern. 
Furthermore, due to the complex space environment \cite{kaushal2016optical}, capacity itself varies over time as ISLs transition between available and unavailable states.
We therefore introduce \textbf{Cap}acity model under a \textbf{D}ynamic \textbf{LSN} (\textbf{Cap-DLSN}), which captures ISL‑availability using semi‑Markov processes, providing a realistic characterization of dynamic capacity variations in LSN. 
On this foundation, we develop the \textbf{M}onte \textbf{C}arlo \textbf{T}hroughput \textbf{E}stimation (\textbf{MCTE}) algorithm, which efficiently samples dynamic traffic distributions and time‑varying capacity constraints to deliver accurate throughput estimates under diverse routing schemes with low computational overhead.
We implement all experimental evaluations on the open-source simulation  platform SNK \cite{snk}, conducting extensive simulations across four operational constellation configurations.

In summary, this paper makes the following contributions:
\begin{itemize}
	\item
	We propose Cap-DLSN, which captures the time-varying availability of ISLs using a semi-Markov process. This model enables realistic characterization of infrastructure capacity under stochastic link failures, a fundamental yet overlooked aspect in prior flow-based analyses.
	\item  We develop MCTE, a Monte Carlo-based method that evaluates achievable throughput under dynamic topology and traffic conditions. 
	Unlike conventional flow-network models that assume optimal routing, MCTE supports any practical routing scheme and reduces computational complexity by  incremental update mechanisms.
	\item  Extensive experiments reveal the quantitative impact of ISL reliability, routing policies, traffic patterns and constellation scale on network throughput, providing actionable engineering guidelines for the design of next-generation LEO satellite networks.
\end{itemize}




\section{Capacity and Throughput in LEO Satellite Networks}
\label{sec:bg}

\subsection{LEO Satellite Network Architecture.}

\begin{figure}[t!]
        \includegraphics[width=1\linewidth]{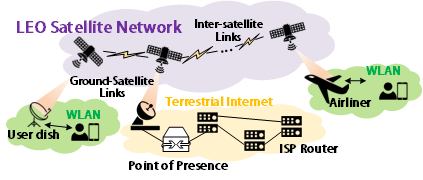}
    \caption{Today's LEO satellite network architecture.} 
        \label{fig:arch}
 \end{figure}

\fig\ref{fig:arch} illustrates the architecture of LEO satellite networks, comprising space, ground, and user segments. 
The space segment consists of thousands of LEO satellites, the ground segment comprises of an operational terrestrial network with ground stations, and the user segment includes of user dishes, terminals and airlines.
The space segment enables global connectivity via high-speed inter-satellite links. 
User terminals first transmit data through multi-hop ISLs\cite{wood2001internetworking,StarlinkTechOverview}, and then transmit it to a ground station (GS) via a ground-to-satellite link (GSL). 
The data subsequently traverses a Point-of-Presence (PoP), where Carrier-Grade NAT (Network Address Translation) performs IP address translation before reaching terrestrial servers.
Terminal users access the Internet through an LSN, with packets initially uplinked to a satellite and then downlinked via a GSL. 
If the serving satellite lacks direct GS connectivity, the data is relayed through multiple ISL hops until it reaches a satellite with GSL access. 
Collectively, by integrating wide-coverage satellite constellation and high-speed communication links,
emerging LSNs promise to provide pervasive, high-throughput and low-latency Internet services globally. 
\subsection{Conventional Capacity and Throughput Quantification in LSN}
\label{sec:model-thp}

The quantification of capacity and throughput in LSNs serves as a core evaluation tool for network design and optimization, providing a quantifiable performance benchmark that directly supports critical engineering decisions such as routing‑scheme development \cite{sun2004routing,liu2022maximum,yang2021maximum}, or topology design \cite{lin2022inter,guo2024constellation,jiang2020reinforcement}.
In most recent works \cite{pachler2021updated,guo2024constellation,basak2023exploring}, network's capacity estimation is usually formulated as a flow network problem, where the maximum flow value $T_{\max}(s)$ obtained through optimization algorithms represents network capacity, and this relationship follows:
\begin{equation}
    \label{eqt:tc1}
    T(s) \leq T_{\max}(s) = C(s)
\end{equation}
where $s$ stands for system that could be nodes (satellites), traffic path or networks.
Depending on the traffic load distribution, network throughput can be analyzed under two scenarios: 1) End-to-end throughput with single source-sink pair and 2) Many-to-many throughput with multiple source-sink pairs.

\noindent\textbf{Scenario 1: End-to-end maximum throughput estimation based on flow network.}
Given a directed graph $\mathcal{G} = (\mathcal{V}, \mathcal{E})$ with a capacity $c(u, v)$ for each edge $e(u, v) \in \mathcal{E}$,
if two nodes in $\mathcal{G}$ are distinguished – one as the source $v_s$ and the other as the sink $v_t$ – then $(\mathcal{G}, c, v_s, v_t)$is called a flow network. The flow $f$ in $\mathcal{G}$ is defined as a function $f: \mathcal{V}\times\mathcal{V} \rightarrow \mathbf{R} $, which meets the following conditions:
\begin{itemize}
    \item Capacity constraints: $0 \leq f(u, v) \leq c(u, v)$. The flow through an edge node must not exceed its capacity.
    \item Flow conservation: For a normal vertex, the sum of incoming flow must equal the sum of outgoing flow.
\end{itemize}

To compute the maximum throughput, the \sn can be modeled as a flow network, where the problem reduces to finding a flow from a source vertex \( v_s \in \mathcal{V} \) to a sink vertex \( v_t \in \mathcal{V} \) that maximizes the flow value \cite{ford1956maximal}, i.e., the classic maximum flow (MF) problem in network flow theory.
In practice, however, we often encounter scenarios with multiple sources and sinks. While this problem can be addressed by introducing a super source (connected to all original sources) and a super sink (connected to all original sinks) \cite{cormen2022introduction}, such an approach may lead to \textit{infinite flow} as the sources and sinks are overlapped.
Consequently, a more accurate and general formulation is to model the problem as a multi-commodity flow (MCF) problem, which explicitly accounts for the distinct flow requirements and interactions among multiple commodities.

\noindent\textbf{Scenario 2: Many-to-many maximum throughput estimation based on multi-commodities flow network.}
The Multi-Commodity Flow (MCF) problem extends classical network flow theory by modeling K distinct flow commodities $(K \geq 2)$, where each commodity $k \in {1,\cdots,K}$ maintains its own source-sink pair $(s_k,t_k)$ and demand $d_k$, subject to shared edge capacity constraints.
The problem also defines $K_i(v_s^i,v_t^i,d^i) \in \{ K_1,K_2,\cdots,K_k\}$ to represent the commodity flowing from source node $v_{src}^i$ to destination node $v_{dst}^i$ with demand $d_i$.
This effectively isolates interference between commodity flows, thus avoiding the occurrence of infinite flows caused by sources-sinks overlapping.
Unlike the single-commodity flow problem, MCF problem defines $f_i(u,v)$ as the proportion of flow $i$ along edge $(u,v)$, where $f_i(u,v) \in [0,1]$. Therefore, the maximum flow problem can be formulated as:

\noindent\textbf{~~~~Objective to}
\begin{eqnarray}
    &\mathop{max} \sum\limits_{i = 1}^{k}d_i& 
\end{eqnarray}
\noindent\textbf{~~~~Subject to}
\begin{eqnarray}
    \label{eq:cst1}
   &  \sum\limits_{i = 1}^{k}   f_i(e)\cdot d_i \leq c(e),\forall e \in \mathcal{E} &\\ 
   \nonumber\\
   \label{eq:cst2}
   &\sum\limits_{(u,v) \in \mathcal{E}}   f_i(u,v) = \sum\limits_{(v,u) \in \mathcal{E}}   f_i(v,u) , \forall i \in \{1,\cdots,k\}&
\end{eqnarray}
where $k$ denotes the number of commodities. \eqt\eqref{eq:cst1} and \eqt\eqref{eq:cst2} represent capacity constraints and flow conservation constraints, respectively.
To address this, the maximum flow problem is broken down into simpler sub-problems, each representing a single commodity flow. These sub-problems can be efficiently solved using classical algorithms such as Ford-Fulkerson \cite{ford1956maximal}, Edmonds-Karp \cite{edmonds1972theoretical}, or Dinic\cite{cormen2022introduction}.

However, the computational complexity of this approach scales linearly with the number of commodities $k$: while single-commodity flow problems exhibit $O(m^2n)$ complexity, their multi-commodity counterparts with $k$ commodities increase to $O(km^2n)$, where $m$ and $n$ denote the number of edges and nodes in the graph, respectively. 
Furthermore, the NP-hard nature of this problem implies that although optimal solutions theoretically exist, they cannot be computed within polynomial time complexity. More critically, the throughput bounds derived from this flow-network-based methodology inherently assume traffic patterns conforming to maximum flow conditions, necessitating the implementation of the network's maximum throughput routing algorithm. 
Nevertheless, practical deployment of such routing algorithms in large-scale, centrally controlled networks presents significant implementation challenges \cite{pan2019opspf,li2024stable}. Consequently, our investigation primarily focuses on evaluating network throughput performance across various routing strategies, which will be comprehensively analyzed in \S\ref{sec:method}.


\subsection{Rethinking the Definition of Capacity and Throughput in LSNs}

For a network communication system, capacity is defined as the maximum amount of data that the system can handle per unit of time, which is only related to the network infrastructure and is a fixed value when the network topology, and facility attributes remain the same \cite{kurose2005computer,peterson2007computer}.
Unlike previous definitions (\eqt\ref{eqt:tc1}), we demonstrate the capacity as the maximum traffic that a network can theoretically handle, which is an inherent characteristic of the network and independent of network policies.  
On the other hand, network throughput is a dynamic value that fluctuates in real time and is affected by various factors such as traffic demand, routing schemes, and even congestion control algorithms.
When these factors are fixed, as the traffic load increases, the throughput gradually increases until it reaches the maximum value $T_{max}$ under the current factors, which is much smaller than the network capacity, i.e.:
\begin{eqnarray}
    \label{eq:tc2}
        T(s;\pi_{\mathcal{N}}) \leq T_{\max}(s;\pi_{\mathcal{N}}) \leq C(s)
\end{eqnarray}
where $s$ stands for the system and could be the satellites, ISLs or networks; 
$\pi_{\mathcal{N}}$ 
represents the set of network policies such as routing schemes, traffic demand, and congestion control.

\section{Capacity Modeling with Dynamic LSN}
\label{sec:method}
%
Owing to the inherent unreliability of ISLs caused by complex space environmental factors, LSNs exhibit fundamental dynamics. This section develops a capacity modeling framework that explicitly accounts for this instability.
\subsection{System model}

\noindent\textbf{LEO Satellite Network.}
To address the temporal variations in the satellite networks,
we denote an ordered time set as $ \mathcal{T}= \{t_1, t_2,\cdots \}$.
The network topology can be considered unchanged between
adjacent time stamps. $t_{i+1}- t_i$ represents the minimum time granularity of scenario change.
Therefore, we model the network topology at each timestamp $t\in\mathcal{T}$ as an undirected graph
 $\mathcal{G}^t = (\mathcal{S},\mathcal{E}^t )$, where $\mathcal{S}$ is the set of network vertices (satellites) and $\mathcal{E}^t$ is the set of undirected edges.
 Due to the inherent unreliability of ISLs in complex space environments, the link set $\mathcal{E}^t$ exhibits time-varying adjacency relationships across different timestamps, resulting in a highly dynamic and dynamic low Earth orbit satellite network.

\noindent\textbf{Traffic Graph.} When traffic is loaded into the LSN $\mathcal{G}$, the link capacity is gradually allocated, and the corresponding traffic path set is represented by the traffic graph $\mathcal{G}_f$, where the edge weights represent the traffic size, and the remaining capacity can be represented by $\mathcal{G}_r$, where the edge weights represent the remaining unallocated capacity of the ISLs. It can be seen that $\mathcal{G}_r = \mathcal{G} - \mathcal{G}_f$.

\noindent\textbf{Traffic demands.}
Given a large number of source GSes $\{u\}$ and destination GSes $\{v\}$, their traffic demands can be expressed by a $|\mathcal{V}_G|\times |\mathcal{V}_G|$ matrix $\mathcal{D}^t$, which can be represented as:
\begin{eqnarray}
    \mathcal{D}^t_{|\mathcal{V}_G|\times |\mathcal{V}_G|} =   
    \begin{bmatrix}
        d^t_{1,1} & 0 & \cdots &d^t_{1,v} \\
        d^t_{2,1}  & 0&\cdots  & 0 \\
        \vdots &\vdots &\ddots &\vdots \\
        0 & 0 & \cdots &d^t_{u,v }
    \end{bmatrix}
\end{eqnarray}
where $d^t_{u,v}$ is the demand bandwidth between source GS $u$ and destination GS $v$ at time $t$, and $|\mathcal{V}_G|$ stands for the number of GSes. We also denote $N_{\mathcal{D}}$ as the number of non-zero elements in $\mathcal{D}$, or the number of traffic demands.

\noindent\textbf{Aggregate throughput.}
For a given traffic demand $d(u,v)$, a traffic path from $p_{u \rightarrow v}$ is generated by the adopted routing scheme in LSN, and the capacity of $p_{u \rightarrow v}$ is defined as the remaining capacity of the bottleneck links it passes through, i.e.,
\begin{eqnarray}
    C_r(p)=& \mathop{min}\limits_{e_i \in p} \{C_r(e_i) \}
\end{eqnarray}
where $C_r(e_i)$ is the residual capacity of ISL $e_i$. 
Assuming that the system allocates resources to each traffic demand as much as possible, the actual allocated bandwidth will not exceed the link capacity, i.e.
\begin{eqnarray}
    T(p) =& \min \{ C_r(p), d(u,v) \}
\end{eqnarray}
Therefore, the aggregate throughput of the entire LSN under the traffic graph $\mathcal{G}_f$ is:
\begin{eqnarray}
    T(\mathcal{G}_f) = \sum\limits_{p \in \mathcal{G}_f} T(p)
\end{eqnarray}

\subsection{Decomposing the LSN Capacity}
\begin{figure}[t!]
    \begin{center}
        \includegraphics[width=1\linewidth]{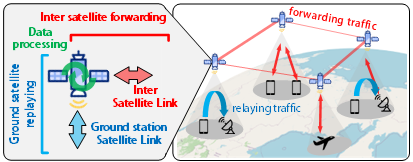}
    \end{center}
    \caption{The ground-satellite \textcolor{blue}{\textbf{relaying traffic}} and inter-satellite \textcolor{red}{\textbf{forwarding traffic}}.} 
        \label{fig:typetraffics}
 \end{figure}

\fig\ref{fig:typetraffics} illustrates the data transmission mechanisms in LSNs, where traffic can be categorized into inter-satellite forwarding traffic and ground-satellite relaying traffic. 
In actual LSN systems such as Starlink\cite{starlink}, thanks to the rapid deployment of ground gateways, the volume of relaying traffic is much larger than that of forwarding traffic. Users relay data packets via satellite to the nearest gateway connected to a Point of Presence (PoP) to achieve Internet connectivity.
ISLs are only used to carry traffic and enable internet access when users cannot relay data packets via gateways, such as in maritime or aircraft scenarios\cite{tanveer2023making}.
Both types of traffic place demands on network capacity.
Compared to relaying traffic, inter-satellite forwarding traffic uses multiple hops to deliver data, which consumes more infrastructure and is therefore more sensitive to infrastructure support capabilities. Furthermore, as business expands and ISL is enabled, this type of traffic will increase, so it is necessary to break down and discuss in detail the capacity issues involved in both types of traffic.

\label{sec:cap}
The capacity in LSN scenario can be categorized into: ISL capacity, satellite capacity, networking capacity and traffic path capacity. The \textbf{ISL capacity} (bandwidth) is defined as the maximum data rate that can be transmitted by an ISL between satellites, which may not be fixed due to free space loss differences caused by the high dynamics of space networks.
On the other hand, \textbf{satellite capacity} refers to the amount of data processed by the satellite per second, which depends on CPU performance and the aggregate capacity of the GSLs.
For example, the Viasat-3\cite{viasat} high-Earth orbit satellites have already reached the capacity with Tbps level.
For \textbf{network capacity}, there are two scenarios depending on how the system works:  satellite-ground relaying (bent-pipe) capable LSN and inter satellite switching/forwarding capable LSN.

\noindent\textbf{Ground satellite relaying capacity.}
In the early stages of deploying Internet satellites in geosynchronous orbit, such as Viasat\cite{viasat}, Chinasat\cite{chinasat}, or LEO constellations such as Starlink GEN1\cite{starlink} and OneWeb\cite{oneweb}, these satellites did not possess the capability for inter-satellite forwarding. 
Instead, they functioned by relaying signals directly from a sender to a receiver using bent-pipe architecture, both of which were simultaneously within the coverage area of an identical satellite.  
Under such ground satellite relaying architecture, the capacity is bottle-necked by the minimum of the onboard data processing ratio and the aggregated uplink/downlink capacity, expressed as:
\begin{eqnarray}
	C_{GSR}(\mathcal{G}) = \sum\limits_{s_i \in \mathcal{V}}  \mathop{min}\{C(s_i), C(e^{gsl}(s_i) + e^{msl}(s_i))\}
\end{eqnarray}

    where $C_{GSR}$ denotes the ground-satellite relaying capacity and $C(s_i)$ is the data processing rate of satellite $s_i$. 
    The aggregated uplink/downlink capacity is modeled through two key link sets of satellite $s_i$: $e^{gsl}(s_i)$ for its ground-station links (i.e., the links between satellites and terrestrial nodes such as user terminals and gateways) and $e^{msl}(s_i)$ for its mobile-station links (e.g., aircraft and remote-sensing satellites).    
This $C_{GSR}$ calculation is also widely used for estimating current LSN system such as Starlink\cite{starlink}.
However, discussing ground-satellite relaying capacity alone may overstate the ability of traffic carrying in entire system, as whose capacity is far smaller due to ISL capacity limitations.

\noindent\textbf{Inter satellite forwarding capacity.}
With the advancement of inter satellite link techniques, mega-constellation networks such as Starlink G2 are increasingly capable of supporting inter satellite forwarding and switching. 
Naturally, our curiosity shifts to determining the maximum traffic that can be handled by the network when datagrams are forwarded at least once through ISLs, i.e., the network capacity of such a system.
This capacity will depend on three key factors: the data processing rate $C(s_i)$, the aggregate uplink and downlink capacity $C(e^{gsl}(s_i))$, and the total capacity of ISLs in the same satellites, denoted as:
\begin{eqnarray}
    \label{eqt:isf}
	&C_{ISF}(\mathcal{G})& = \\
    \nonumber
    &&\sum\limits_{s_i \in \mathcal{V}}  \mathop{min}\{C(s_i), C(e^{gsl}(s_i)), \sum\limits_{e_j \in \mathcal{E}(s_i)} C(e_j)\}\\
    \nonumber
    &&=\sum\limits_{e_j \in \mathcal{E}} C(e_j)
\end{eqnarray}
where $C_{ISF}$ is inter satellite forwarding capacity and $\mathcal{E}(s_i)$ represents the ISL set that adjacent to $s_i$. 
For ISL-capable LSNs, estimating their ISF capacity better reflects their network's ability to carry traffic, thereby better quantifying the LSN system's capabilities when faced with global traffic. 

This requires better modeling of ISF capacity. In the following discussion, we will not consider GSR capacity, but will focus on ISF capacity, which is constrained by ISL capacity according to \eqt\ref{eqt:isf}.

\subsection{Understanding the unreliable ISLs}
\label{sec:uISL}
Due to the highly dynamic nature of satellite networks and the unpredictable space environment effects, inter-satellite links exhibit inherent non-reliability and may experience intermittent disruptions caused by dynamic relative motion \cite{lee2021connectivity}, Sun outage \cite{choudhary2024inter}, or active attacks \cite{giuliari2021icarus}, ultimately degrading network capacity.
Given the superior performance characteristics of free-space optical communication, we adopt the assumption that all satellites employ optical ISLs and accordingly develop a network capacity model accounting for the unreliable nature of optical links in free space.
Under this framework, each optical ISL $e$ is subject to random interruptions that demonstrate memory-less properties with constant failure rates.
Consequently, both the available time of the ISL $X$ and the interruption duration $Y$ can be modeled as an exponential distribution \cite{wu2013analysis} \footnote{The study \cite{wu2013analysis} models ISL interruptions as independent Bernoulli events. We simplify the parameterization by treating the link‑available time as a continuous random variable and modeling it using an exponential distribution.}.
 When ISL failures occur, the onboard Acquisition, Tracking, and Pointing (ATP) system \cite{kaymak2018survey} initiates realignment procedures through sequential acquisition, tracking, and pointing operations.
The recovery duration $Y$ is assumed to follow a log-normal distribution due to the compound effects of pointing inaccuracies, attitude control errors, and the inherent right-skewed distribution of system restoration times. 
coherently, the complete statistical characterization is given by:
\begin{eqnarray}
    X_e \sim Exp(\lambda),~ Y_e \sim LogN(\mu,\sigma^2)
\end{eqnarray}
As $X_e$ and $Y_e$ occur in alternating renewal sequences of ISL $e$, and assuming they are independent and identically distributed (i.i.d.), the availability of ISLs can be described by an semi-Markov process\cite{Gallager2013Stochastic}. As is illustrated in \fig\ref{fig:arp}, the availability of ISL $Z_{e}$ can be considered as the reward function which is formulated as:
\begin{figure}[t!]
        \includegraphics[width=0.85\linewidth]{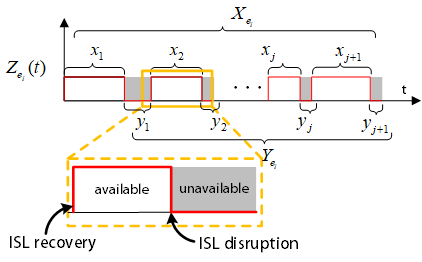}
    \caption{Alternating renewal process in ISL maintenance. The red line with high (1) and low (0) status denotes ISL availability function $Z_e(t)$.} 
        \label{fig:arp}
 \end{figure}

\begin{equation}
    Z_{e}(t) = \left\{
    \begin{aligned}
    1,~ &\text{if}~  \sum\limits_{j = 1}^{n} (x_j+y_j) < t <\sum\limits_{j = 1}^{n} (x_j+y_j) +y_{j+1} \\
    0,~&\text{if}~  \sum\limits_{j = 1}^{n} (x_j+y_j) +y_{j+1}< t <\sum\limits_{j = 1}^{n+1} (x_j+y_j) \\
    \end{aligned}
    \right.
\end{equation}
where $1<j<n$ and $x_j+y_j$ is the $j$-th cycle of renewal process. If $Z_{e}(t)=1$, the ISL is available for data transmission, otherwise the ISL is unavailable. 
Then the expectation of $ E[Z_{e}(t)] $ can be formulated as:
\begin{eqnarray}
    E[Z_{e}(t)] = \dfrac{E(X_{e})}{E(X_{e}) + E(Y_{e})}
\end{eqnarray}
\textbf{Proof:}
Define the $j$-th renewal cycle $h_j =x_j+y_j$, with the expectation $E[H]=E[X]+E[Y]$.
The reward $Z(t)=1$ if the system is "On" at time $t$, else $Z(t)=0$. 
By the renewal-reward theorem\cite{Gallager2013Stochastic}, the long-term time proportion is the expected reward per cycle divided by the expected cycle length:
\begin{eqnarray}
    \nonumber
    E[Z(t)] = \lim\limits_{t \to \infty}  \dfrac{1}{t}\int_{0}^{t} Z(\tau) \,d\tau \\
     =  \dfrac{E[X]}{E[S]} = \dfrac{E(X)}{E(X) + E(Y)}
\end{eqnarray}
where $X$ and $Y$ can be arbitrarily distributed, as long as they are independently and identically distributed. The above conclusion holds true regardless of whether $X,Y$ are correlated.
Based on the network capacity model with bottleneck ISL (\eqt\ref{eqt:isf}), the network capacity with unreliable ISL is:
\begin{eqnarray}
   \label{eq:netcap2}
   C(\mathcal{G}^t) = \sum\limits_{e \in \mathcal{E}}Z_{e}(t) \cdot C(e)
\end{eqnarray}
Thus, the expectation of network capacity can be formulated as follows:
\begin{eqnarray}
\label{eqt:exp_cap}
	\nonumber
	E[C(\mathcal{G}^t) ] &= &\sum\limits_{e \in \mathcal{E}} C(e) E[Z_{e}(t)]\\
	&=& \sum\limits_{e \in \mathcal{E}} C(e) \frac{  1 }{1 + \gamma\lambda}
\end{eqnarray}
where $1/\lambda$ represents the expected value of $X$ and $\gamma = e^{\mu+\sigma^2/2}$ denotes the expected value of $Y$ \cite{Gallager2013Stochastic}. 
To simplify our model and focus more on the analysis above the physical layer, we treat both $\lambda$ and $\gamma$ as configurable parameters in the model.

\section{Monte Carlo Throughput Estimation in Dynamic LSNs}

 \begin{algorithm}[t!]
    \caption{ Monte Carlo Throughput Estimation under Deterministic Topology (\textbf{MCTE-DT}) }
    \label{alg:mcte-dt}
    \LinesNumbered 
    \KwIn{ LSN graph $\mathcal{G}$ with ISL capacity at each timestamp}
    \KwOut{ Aggregate throughput $T{(\mathcal{G}_f)}$, traffic graph $\mathcal{G}_f$}

		$\mathcal{D}^t\gets$\texttt{TrafficGen()}\textup{//traffic generation} \\
        \While{  $d_{u,v}!=0 ,d_{u,v}\ \in \mathcal{D}^t $ }{
        
			$p_{u \rightarrow v}\gets$\texttt{\textup{path($u,v;\mathcal{G}$)}};~\textup{//routing}\\
            $T(p_{u\rightarrow v}) \gets min \{C_r(p_{u\rightarrow v}),d_{u,v} \}$;\\
			\If{ $T(p_{u\rightarrow v})!=0$}{
				\textup{//update traffic graph} \\
				$\mathcal{G}_f$.\texttt{\textup{add($p_{u \rightarrow v},T(p_{u\rightarrow v})$)}};\\
				$\mathcal{G}_r \gets \mathcal{G} - \mathcal{G}_f$;
				$T_{\mathcal{G}_f^t} \gets  T_{\mathcal{G}_f} + T(p_{u\rightarrow v}) $;
			}
            
        }
	\Return{$T{(\mathcal{G}_f)},\mathcal{G}_f$}
    \end{algorithm}



%
Through the proposed Cap-DLSN model, we establish a framework for determining real-time ISL capacity across temporal variations, thereby facilitating precise throughput estimation in dynamic network conditions.
The throughput and its upper bound of LSN is affected by various factors such as routing and traffic schemes. In order to reflect the performance of these schemes on the network, routing and traffic schemes should be dynamically adjusted independently of the throughput estimation process.
Therefore, we propose a \textbf{M}onte \textbf{C}arlo \textbf{T}hroughput \textbf{E}stimation (\textbf{MCTE}) method that statistically analyzes changes in throughput under complex conditions by inputting synthetic traffic into an dynamic LSN.
The two complementary algorithms are proposed for throughput estimation: 
\begin{itemize}
    \item \textbf{MCTE-DT:} Monte Carlo throughput estimation under deterministic topology.
    \item  \textbf{MCTE-ST:} Monte Carlo throughput estimation under stochastic topology.
\end{itemize}
The MCTE-DT algorithm is particularly suitable for offline network scenarios with accurate topological snapshots, as it can accurately estimate the throughput at each moment. However, this is difficult to achieve in practice due to the large scale and high dynamism of LSNs. Therefore, MCTE-ST integrates a Semi-Markov Process model for ISLs, obtains the current topological state through state transitions, and achieves throughput estimation with limited error and higher efficiency.

\subsection{Throughput Estimation under Deterministic Topology}
Assuming that the topology of LSN is deterministic at each timestamp, the aggregate throughput of the LSN over a period can be obtained by iterating through each timestamp and accumulating the successfully allocated bandwidth of each generated traffic paths. The main procedure of MCTE-DT ares shown in \alg\ref{alg:mcte-dt}.


The \texttt{TrafficGen()} function generates a traffic demand matrix following a specific statistical distribution (line~1). 
The estimation process supports arbitrary routing algorithms without being constrained by maximum flow routing (line~3). 
When successful routing with allocated bandwidth occurs (line~4), end-to-end data transmission proceeds according to the assigned bandwidth, with actual throughput being progressively accumulated (lines~5--10).

However, this approach necessitates a global view of the LSN topology and requires recomputing the traffic graph $\mathcal{G}_f$ at each timestamp. 
This leads to unstable routing decisions for each traffic demands and consequently produces fluctuating throughput measurements. 
In practice, network paths typically remain stable over extended time periods. 
Moreover, the computational complexity becomes significant when recalculating the traffic graph at each time step: 
Given a routing complexity of $\mathcal{O}(N_{\mathcal{V}}^2)$, the overall complexity of MCTE-DT becomes $\mathcal{O}(|\mathcal{T}|N_{\mathcal{D}}N_{\mathcal{V}}^2)$, where $|\mathcal{T}|$ represents the number of timestamps and $N_{\mathcal{D}}$ denotes the number of traffic demands.
To address these limitations, we exploit the temporal consistency properties of traffic graph $G$ and propose the MCTE framework under Stochastic Topology.



\subsection{Throughput Estimation under Stochastic Topology.}

 \begin{algorithm}[t!]
    \caption{ Monte Carlo Throughput Estimation under Stochastic Topology (\textbf{MCTE-ST})}
    \label{alg:mcte-st}
	\LinesNumbered 
    \KwIn{ 
	 traffic graph $\mathcal{G}^{t-1}_f$ , traffic paths $\mathcal{P}^{t-1}$  and ISL availability indicator $Z_e(t-1)$
	 }
    \KwOut{Aggragate throughput $T(\mathcal{G}^t_f)$,traffic graph $\mathcal{G}^t_f$ }



	\SetKwFunction{fgu}{fupdate}
    \SetKwProg{Fn}{def}{\string :}{}
	
	\tcp{Stage 1. initialization and state transition}
	$\Delta\mathcal{D}\gets \mathbf{0}_{N_D \times N_D}$, $\Delta\mathcal{P}\gets \emptyset$;\\
	$Z_e(t) \gets $\texttt{state\_trans($Z_e(t-1),P_e$)};\\

		$Z_e^\prime(t) \gets Z_e(t) - Z_e(t-1)$;\\

		\tcp{Stage 2. path collection}
		$\mathcal{D}^t \gets $ \texttt{TrafficGen()};\\
			\For{$z^\prime_e(t)  \in Z_e^\prime(t)$ }{ 


				\uIf{($\exists p \in \mathcal{G}_f^{t-1}, e_{ij} \in p$ \textbf{and} $z^\prime_e(t)=-1$ ) 
				}{
					\tcp{ISL interuption}
					$\Delta\mathcal{D}[u_p,v_p] \gets \mathcal{D}^t[u_p,v_p]$;
					$\Delta\mathcal{P}$.\texttt{\textup{add($p$)}};\\

				}
				\uElseIf{$\exists p \in \mathcal{G}_f^{t-1}, e_{ij} \notin p$  \textbf{and} $i_e,j_e \in p$ 
				\textbf{and} $z^\prime_e(t)=-1$}{\tcp{ISL recovery}
				$\Delta\mathcal{D}[u_p,v_p] \gets \mathcal{D}^t[u_p,v_p]$;
				$\Delta\mathcal{P}$.\texttt{\textup{add($p$)}};\\

				}
				\ElseIf{$\exists p \notin \mathcal{G}_f^{t-1} $ \textbf{and} $\mathcal{D}^t[u_p,v_p] $}{
					\tcp{new traffic}
					$\Delta\mathcal{D}[u_p,v_p] \gets \mathcal{D}^t[u_p,v_p]$;
				$\Delta\mathcal{P}$.\texttt{\textup{add($p$)}};\\
				}


		}
		\tcp{Stage 3. traffic flow update}

		\For{$p \in \Delta\mathcal{P}$ \textbf{and  $ d_{u,v} \gets  \Delta\mathcal{D}[u_p,v_p]$}}{
			
			$T^\prime \gets $\fgu($\mathcal{G}_f^{t-1},p,u_p,v_p,d_{u,v}$);
		
		}
		$T(\mathcal{G}^{t}_f) \gets T^\prime$;
      
	\Return{$T(\mathcal{G}^t_f), \mathcal{G}^t_f$ }\\
	\tcp{traffic flow update function}
	\Fn{\fgu{ $\mathcal{G}_f^{t-1},p^{t-1},u,v,d_{u,v}$} }{
		$\mathcal{G}^{t-1}_f \gets \mathcal{G}^{t-1}_f$.\texttt{\textup{del($p^{t-1}_{u \rightarrow v},T( p^{t-1})$)}};\\
		$p^t \gets$	\texttt{\textup{path($u,v;\mathcal{G}^t$)}};\\
		$T(p^t) \gets min \{C_r(p^t),d_{u,v} \}$;\\
		$\mathcal{G}^{t-1}_f \gets  \mathcal{G}^{t-1}_f$.\texttt{\textup{add($p^t,T(p^t)$)}};\\
		$T(\mathcal{G}^{t-1}_f) \gets T(\mathcal{G}^{t-1}_f)  - T( p^{t-1}) +T( p^{t}) $;\\
	}
	\Return{$T(\mathcal{G}^{t-1}_f)$ }\\
    \end{algorithm}

In the CAP‑uLSN model, the state of each ISL follows an alternating renewal process, leading to stochastic variations in the network topology. However, in practice, traffic fluctuations and path changes tend to remain stable over short timescales under typical routing schemes. To estimate throughput under these conditions, we propose the Monte Carlo Throughput Estimation under Stochastic Topology (MCTE‑ST) method.
In the MCTE-ST algorithm, the parameters (such as $\lambda$ and $\gamma$) of the ISL state transition probability (\eqt\ref{eqt:exp_cap}) are derived from the statistical modeling of ISL availability in the Cap-DLSN model.
This temporal consistency allows us to significantly reduce the complexity of throughput estimation while producing more stable routing paths, thereby ensuring that the derived paths closely match actual routing scenarios and providing more realistic throughput estimates.

As established in \S~\ref{sec:method}.\ref{sec:uISL}, the conditional probability distribution of each ISL's future state depends solely on its current state, confirming the Markov property of ISL availability. 
Moreover, since ISL disruptions and recoveries follow distinct probability distributions, the state transitions constitute a semi-Markov process. 
The corresponding state transition matrix is given by:
\begin{equation}
    P_e = \begin{bmatrix}
        1/\lambda & 1-1/\lambda \\
        1-\gamma  & \gamma
    \end{bmatrix}
\end{equation}
where $1/\lambda$ represents the expected duration of ISL availability and $\gamma$ denotes the probability of remaining in the unavailable state.


Therefore, the throughput estimation method leveraging the Markovian properties of ISLs is presented in Algorithm~\ref{alg:mcte-st}, which comprises three sequential stages:

\noindent\textbf{Stage 1):} During throughput computation at each timestamp $t$, ISL state transitions are derived from previous states using the \texttt{state\_trans()} function (line~2) under the semi-Markov process framework to determine current topology adjacency, with forward differencing employed to detect ISL state changes.

\noindent\textbf{Stage 2):} Traffic paths affected by ISL state transitions are identified and processed accordingly:
If an ISL becomes interrupted, all associated traffic paths are offloaded, their throughput recomputed, and the traffic graph $\mathcal{G}_f$ updated;
If an ISL is restored, throughput is recalculated for all traffic paths involving the satellites at both ends of the ISL, followed by traffic graph $\mathcal{G}_f$ updates;
If the ISL state remains unchanged, no action is required.

\noindent\textbf{Stage 3):} Incremental updates are performed to maintain consistency between network dynamics and both the traffic graph $\mathcal{G}_f$ and throughput $T(\mathcal{G}_f)$.  
Lines~19--25 detail the traffic graph update procedure:
First, remove traffic paths affected by ISL state changes from the traffic graph;
Then, reroute traffic demands between source and destination using the tested routing scheme;
Next, allocate throughput based on remaining bandwidth capacity;
Finally, update the throughput for the entire traffic graph.

\begin{figure}[t!]
        \includegraphics[width=1\linewidth]{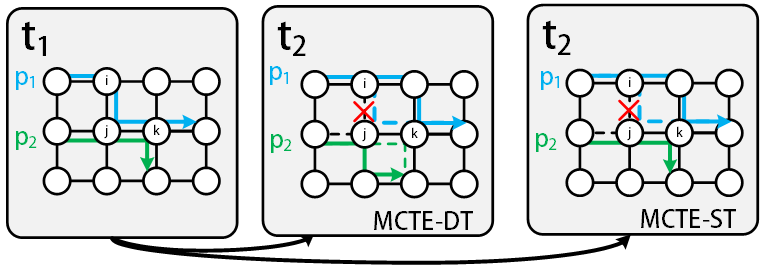}
    \caption{ 
        Comparison of MCTE-DT and MCTE-ST methods during ISL $e_{ij}$ failure events at time $t_2$.
    While both MCTE methods recompute $p_1$'s traffic path, MCTE-DT additionally recomputes $p_2$'s route, implicitly assuming that the routing scheme is unstable and sensitive to topology changes. 
    In contrast, MCTE-ST only updates paths affected by the failed ISL ($p_1$ ), maintaining unaffected flows ($p_2$) with their previous routing decisions. } 
        \label{fig:mcte}
 \end{figure}
 
The complexity of MCTE-ST is primarily determined by the affected traffic path collection in stage 2 and the traffic update operation in stage 3.
If a routing scheme with $O(N_{\mathcal{V}}^2)$ complexity is used, the complexity within the $\mathcal{T}$ time is $O(|\mathcal{T}|KN_{\mathcal{V}}^2)$, where $ K = \mathop{min}(N_{\Delta \mathcal{D}}, |\Delta \mathcal{P}|)$, $N_{\Delta \mathcal{D}}$ denotes the number of affected traffic demands and  $|\Delta \mathcal{P}|$ represents the number of affected traffic paths.
However, due to the temporal consistency in traffic paths under most routing schemes,
only when ISL availability changes occur should the relevant routes be recalculated,
that is, $K< N_{\Delta\mathcal{D}}<< N_{\mathcal{D}}$, resulting in an extremely limited number of routing scheme recalculating, far fewer than the MCTE-DT algorithm.
The main differences between the two methods are shown in \fig\ref{fig:mcte}.

\section{Simulation}


In this section, we evaluate the capacity and throughput of current LSN systems through experiments, using our proposed dynamic capacity model (Cap-DLSN) and Monte Carlo Throughput Estimation (MCTE) method.
This study first adjusts the parameters of the Cap-DLSN model and conducts a network capacity sensitivity analysis; 
Subsequently, in an dynamic network environment, the proposed MCTE methods are compared with traditional flow-based network methods to validate their performance advantages; 
furthermore, by applying the MCTE under different routing schemes and constellation configurations, the impact of routing strategies on network aggregate throughput is thoroughly investigated; 
finally, by adjusting traffic patterns, the influence patterns of traffic modes on network performance are systematically analyzed.

\begin{figure}[t!]
\centering
	    \subfigure[Constellation networks]
    {   
        \begin{minipage}{0.48\textwidth}
        \includegraphics[width=\linewidth]{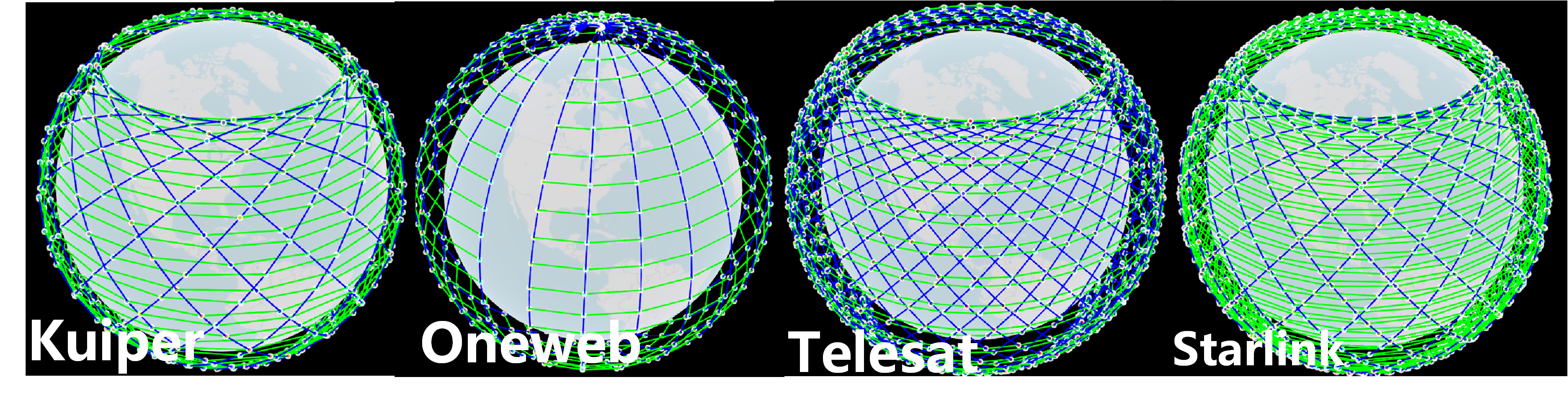}
        \end{minipage}
    }   \subfigure[Ground stations and population distribution\cite{ciesin_gpwv4} for traffic pattern generation.]
    {   
        \begin{minipage}{0.48\textwidth}
        \includegraphics[width=\linewidth]{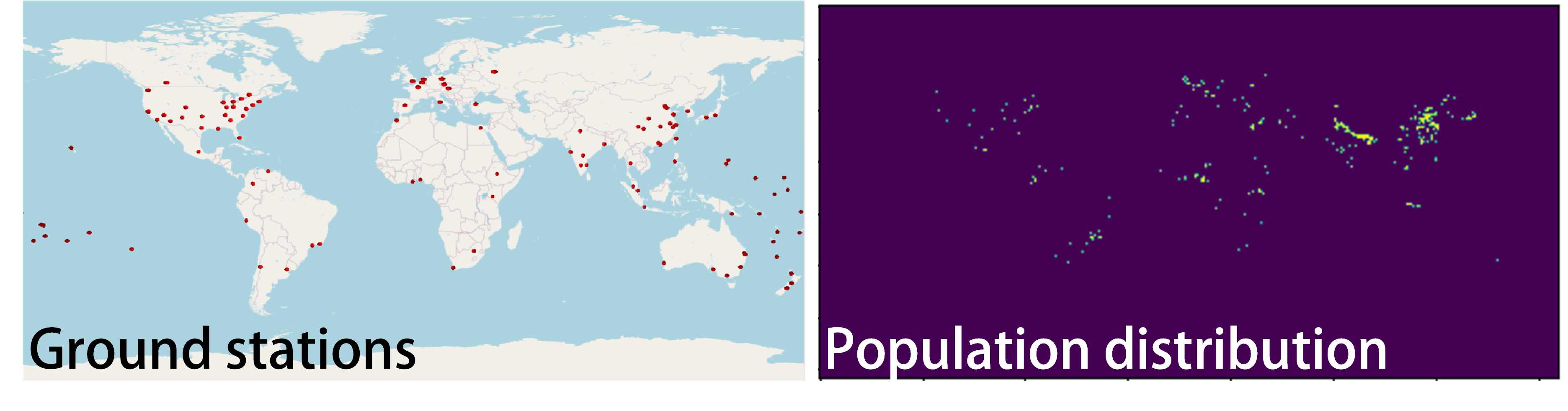}
        \end{minipage}
    }
\caption{Simulation scenarios.}
\label{fig:sce} 
\end{figure}

\begin{table}[t!]
	\caption{Parameters setting for evaluation across all constellations.}
	\centering
  \scalebox{0.9}{
		  \begin{tabular}{l|ccccc}
			\toprule[1pt]
			   &Year& Planes   & Sat/Plane    &Altitude&Inclination   \\  \toprule[1pt]
			Kuiper\cite{kuiper}&2019& 17&34&630 km&$51.9^\circ$ \\
			OneWeb\cite{oneweb}&2017& 12    & 49&1200 km&$87.9^\circ$ \\
			Telesat\cite{telesat}&2020& 40   & 33 &1325 km &$50.9^\circ$\\
			Starlink\cite{starlink}&2020& 22    &72&550 km&$53^\circ$ \\
			\toprule[1pt]
		\end{tabular}
		  }
	\label{tab:cons-param}
\end{table}

\subsection{Experiments setups}

\noindent\textbf{Constellation setting.} 
\fig\ref{fig:sce} (a) depicts the four public LEO satellite networks evaluated in this study: OneWeb \cite{oneweb}, Kuiper \cite{kuiper}, Telesat \cite{telesat}, and Starlink \cite{starlink}. The \tab\ref{tab:cons-param} summarizes their key configuration parameters. We assume that each constellation supports inter-satellite data relay via ISLs, all of which adopt the +Grid topology\cite{bhattacherjee2019network}.

\noindent\textbf{Traffic patterns.} 
\fig\ref{fig:sce} (b) illustrates the global ground-station distribution and population density used for traffic generation. The traffic demand matrix $\mathcal{D}$ is constructed through a two-step process. First, the source and destination endpoints are selected from more than 100 ground stations worldwide, with a selection probability proportional to the density of the local population \cite{bhattacherjee2019network}. Second, these endpoints are paired to form traffic flows using either a random pairing scheme or a gravity model \cite{nucci2005problem}, in which the probability of pairing decays as a function of the distance between stations.
The request data rate of each traffic demand is scaled in $0 \sim 5$ Gbps and proportional to the population products of the GSes pairs. For each simulation, over 1000 end-to-end loads are generated for each timestamp at 10 second intervals over a 3-hour period.

\begin{table}[t!]
	\caption{Parameters setting for evaluation across all constellations.}
	\centering
\renewcommand{\arraystretch}{1.1}  
	\scalebox{0.9}{
		\begin{tabular}{lll}
			\toprule[2pt]
			Parameter                               & Symbol   & Setting         \\  \toprule[1pt]
			GSL Aggregate Capacity                & $C^{gsl}$    & 100 Gbps \\
			ISL Capacity                & $C(e)$    & 10 Gbps \\
			log-mean of ISL recovery duration             & $\mu$    & $[0\sim 10]$   \\
			log-deviation of ISL recovery duration             & $\sigma$    & $[0\sim 4]$   \\
			ISL disruption probability & $\lambda$    & $[10^{-6}\sim 10^{-3}]$  \\
			simulation time && 3 hours\\
			each traffic demand data rate& &$[0\sim5]$ Gbps\\
			\toprule[2pt]\\
		\end{tabular}
	}
	\label{tab:param}
\end{table}

\noindent\textbf{Routing schemes.} 
To provide a comprehensive benchmark for throughput evaluation under various routing strategies, we analyze three representative schemes: shortest-distance path (SDP) routing \cite{chen2024shortest}, geographic routing (GEOR) \cite{henderson2000distributed}, which are well-established in the literature; and minimum-load path (MLP) routing \cite{liu2020load}, a strategy commonly implemented in operational LEO satellite constellations for dynamic traffic management.

\noindent\textbf{Throughput estimation algorithms.}
We evaluate the 1-commodity Maximum Flow (MF), Multi-Commodity Flow (MCF), and our proposed Monte Carlo Throughput Estimation under Deterministic Topology (MCTE-DT) and Stochastic Topology (MCTE-ST) under a unified traffic demand matrix and routing scheme.

\noindent\textbf{Parameters setting.}
\tab\ref{tab:param} shows the detailed parameters setting in simulation. The ISL capacity is set to 10 Gbps, aligning with the demonstrated capabilities of current-generation optical inter-satellite links \cite{choudhary2024inter}. 
The aggregate GSL capacity per satellite is set to 100 Gbps, a conservative estimate based on the aggregate bandwidth of multiple user and gateway beams per satellite \cite{choudhary2024inter} 
\footnote{based on the reference\cite{choudhary2024inter}, 
the GSL and ISL capacity parameters that we have set are fully achievable in orbit}. 
The onboard data processing rate $C(s_i)$ is implicitly assumed to be non-bottlenecking (i.e. greater than the sum of its incident link capacities) for these experiments, focusing the analysis on link-capacity constraints.
The ISL failure rate $\lambda$ and recovery duration parameters $(\mu, \sigma)$ are varied in various ranges to perform the sensitivity analysis in \Sec\ref{exp:sens}. 
The baseline values for comparative throughput evaluation in \Sec\ref{exp:comp} are chosen to maintain a long-term ISL availability $E[Z_e(t)]$ of approximately 80\% (e.g., $1/\lambda =  10^4$, $\gamma =  4\times 10^2$), representing a scenario with non-persistent but recurring link disruptions.

\noindent\textbf{Implementation.}
We utilize the open-source SNK simulator \cite{snk} to implement the Cap-DLSN and MCTE algorithm in its \textit{Analyzer} kit and build simulation scenarios based on the above constellations, which are shown in \fig\ref{fig:sce}.
The experiments settings are shown in \tab\ref{tab:param}.
Simulations were run on a PC with Ubuntu 18.04.2 LTS (64 bit), an Intel Core i7-10700 at CPU 2.9 GHz $\times 16$, and 32 GB RAM.

\subsection{Sensitivity Analysis of Cap-DLSN Model}
\label{exp:sens}

\begin{figure}[t!]
	    \subfigure[ Capacity vs ISL availability duration $(1/\lambda)$ at $\gamma=10^2$.]
    {   
        \begin{minipage}{0.45\textwidth}
        \includegraphics[width=\linewidth]{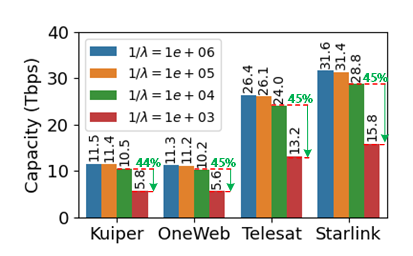}
        \end{minipage}
    }
	
	\subfigure[Capacity retention under ISL recovery time distribution.
	]
    {
    \begin{minipage}{0.45\textwidth}
    \includegraphics[width=\linewidth]{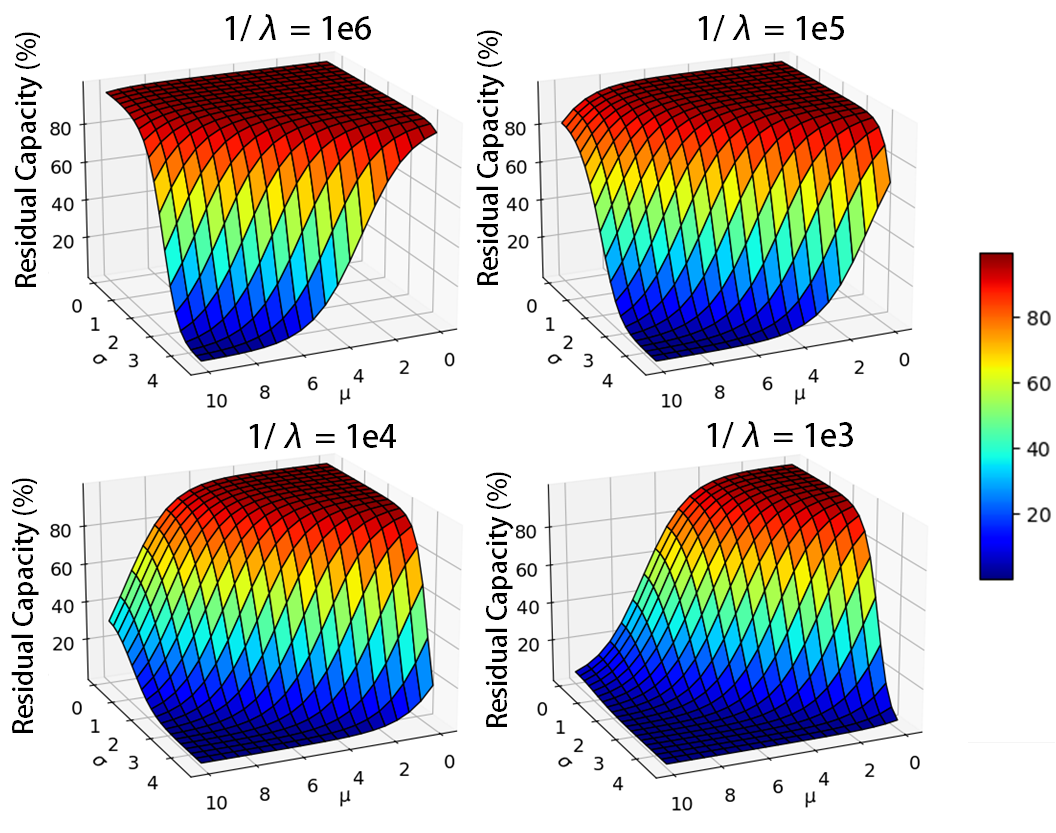}
    \end{minipage}
    }
    \caption{Capacity sensitivity analysis under unreliable ISL conditions.} 
       \label{fig:cap}
   \vspace{-1em}

   \end{figure}




\fig\ref{fig:cap} (a) 
illustrates the LSN capacity distribution based on different ISL maintenance durations, with $ \gamma = 1 \times 10^2$, representing the average ISL recovery duration in seconds.
 The results show a clear trend: as $1/\lambda$ decreases, the network capacity drops significantly.
 For example, when $1/\lambda = 1 \times 10^4$, the capacity drops to $90\%$ of the initial value.However, when $1/\lambda$  is reduced to $ 1 \times 10^3$, the capacity sharply falls to below 60\%. 
 These findings highlight the crucial importance of minimizing ISL disruptions to maintain a high level of network capacity, especially in dynamic satellite environments.

\fig\ref{fig:cap} (b) also compares the impact of average ISL available duration on capacity performance under different fault parameter settings. 
Analysis reveals that as the average ISL available duration decreases, stricter fault time parameters (i.e., smaller $\mu$ and $\sigma$) are required to maintain a suitable network capacity. 
Specifically, when $1/\lambda = 1 \times 10^4$ (corresponding to an average ISL available duration exceeding 2 hours), the summary recovery duration window of ISL must be controlled within $403 \pm 7.3$ seconds ($\mu \leq 6, \sigma \leq 2$).
This sensitivity characteristic is consistent with the random process theory prediction in Equation \eqt\ref{eq:netcap2}: when $E[Z_e(t)] = 1/(1+\gamma\lambda) < 0.9$, the network enters a high-loss state. 
For $\gamma = 10^2$, the critical point $\lambda = 10^3$ corresponds exactly to the inflection point in the figure. 
In actual constellation operations, the system should be operated in the region where $E[Z_e(t)] > 0.9$ by increasing $\lambda$ (reducing the failure rate) or decreasing $\gamma$ (accelerating recovery).

\subsection{Comparison of Throughput Estimation Methods}
\label{exp:comp}

\begin{figure}[t!] 
    \centering  

        \subfigure[Load-throughput curves. The term "load traffics" refers to the total number of individual traffic requests generated from the demand matrix $\mathcal{D}$. The grey dash line stands for the LSN capacity while the black dash line stands for the aggregate throughput expectation ($E(||\mathcal{D}||_1)$).]
        {
            \begin{minipage}{.46\textwidth} 
                \includegraphics[width=1\linewidth]{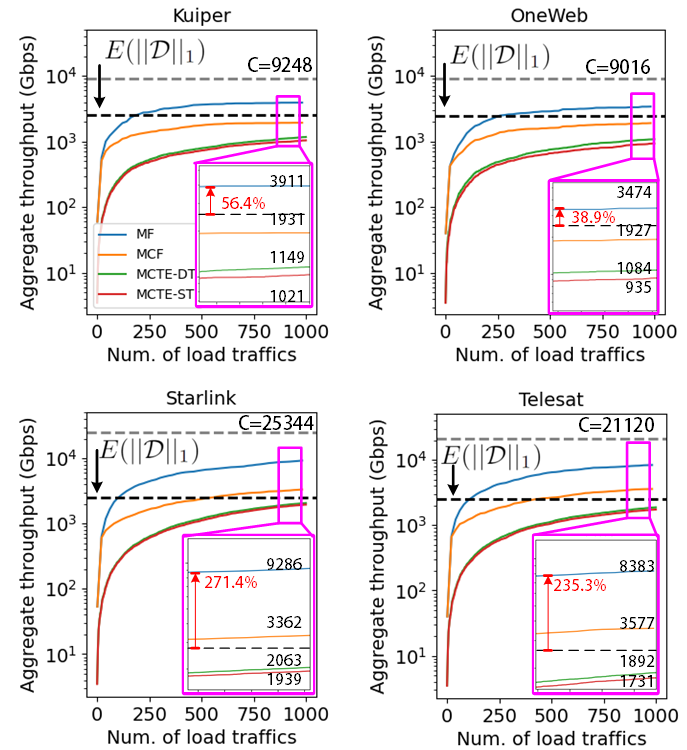} 
				\vspace{0.02\baselineskip} %
			\end{minipage}
        }
		\centering

		\subfigure[Avg. time consumption of throughput estimation at single timestamp.]
        {
\renewcommand{\arraystretch}{1.1}  
            \scalebox{1}{
                \begin{tabular}{lcccc}
					\toprule[1pt]
					 & MF &MCF &MCTE-DT &MCTE-ST  \\ 
					\midrule
					Time (sec) & 25.35 & 55.66 & 1.22 & \textbf{0.16}   \\
					Std (sec) & $\pm 1.07$ & $\pm 1.02$ & $\pm 0.45$ & $\pm 0.07$  \\
					\bottomrule[1pt]
			\vspace{1pt} %
					\end{tabular}
            }
        }

    \caption{Load-throughput curves and time consumption of throughput estimation methods.} 
        \label{fig:load-thp}
\end{figure}

\fig\ref{fig:load-thp} (a) presents a comparative analysis of different throughput estimation methods under the same traffic and routing scheme in four public satellite constellations.
All constellations employ the same unreliable ISL model, with network availability controlled at approximately 80\% via parameters, and dashed lines indicating the upper limit of theoretical capacity. 
Each injected traffic request is uniformly distributed between $[0\sim 5]$ Gbps, with a total of 1,000 load traffics, resulting in an expected total request volume $E (||\mathcal{D}||_1) = 2.5$ Tbps, where $||\mathcal{D}||_1$ represent the sum of requiring bandwidth of among all traffic demands, as indicated by the black dotted line in the \fig\ref{fig:load-thp}.
We observe that 
MF algorithms demonstrate the highest aggregate throughput, exceeding at least over 38.9\% (OneWeb) throughput compared with the expectation across all four constellations.
In contrast, by representing different traffic flows through multiple commodity flows, MCF isolates the infinite flow situation caused by overlapping source and destination points, significantly reducing overestimated throughput and keeping network throughput lower than $E (||\mathcal{D}||_1)$ in Starlink and Telesat, but still overestimated in Kuiper and OneWeb.
The total throughput of the MCTE-DT and MCTE-ST methods we proposed was the lowest. Under maximum load conditions, the network throughput remained below expectations $E (||\mathcal{D}||_1)$.
This is because when allocating bandwidth, there may be insufficient capacity in the traffic path, resulting in allocations below the requested amount, which causes the overall throughput to be lower than the expected value of traffic $\mathcal{D}$.

Compared with traditional flow network based methods, the MCTE method more accurately evaluates network throughput by injecting synthetic traffic, rather than obtaining maximum flow through integer programming and other algorithms and simply equating it with network traffic.
This avoids overestimating throughput and indirectly reflects that the throughput of LSN networks may be far below its capacity.
In terms of computational efficiency, the MCTE method requires less than 5\% of the computational time to complete throughput estimation compared with flow network based methods, as shown in \fig\ref{fig:load-thp} (b).
In addition, under the semi-Markov process model, MCTE-ST only recalculates the throughput corresponding to new traffic paths or traffic paths with ISL state transitions. 
In contrast, MCTE-DT recomputes the complete traffic graph
$\mathcal{G}_f$ at each moment, which affects the temporal consistency of the traffic paths and causes differences in the results compared to MCTE-ST and consumes more time $7\times$ than MCTE-ST.

\begin{figure}[t!]
    \begin{center}
        \includegraphics[width=0.85\linewidth]{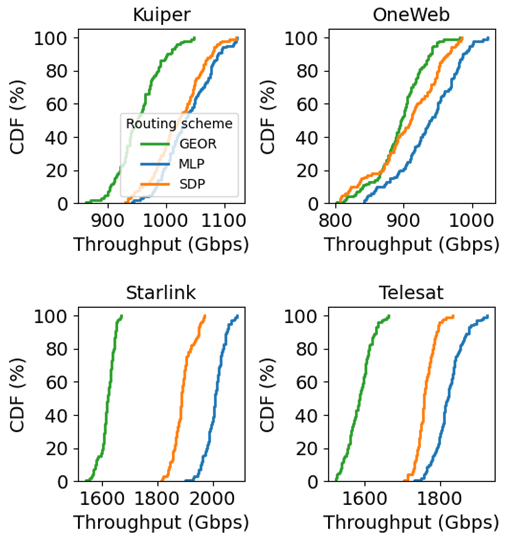}
    \end{center}
    \caption{Throughput distribution of different routing schemes using MCTE-ST.} 
        \label{fig:route-thp}
 \end{figure}

\subsection{Impact of Routing Scheme on Aggregate Throughput}

\fig\ref{fig:route-thp} evaluates three representative routing schemes in four constellations. 
Most flow-network-based methods inherently assume maximum throughput path routing, which limits their ability to evaluate arbitrary routing schemes. 
Thus, this evaluation is conducted solely using the MCTE method.
We observe that the MLP algorithm achieves superior overall throughput by consistently selecting paths with maximum residual capacity in the global network view. 
Similarly, the SDP routing scheme minimizes ISL occupancy by achieving the shortest traffic path length, thereby improving aggregate throughput .
In contrast, while GEOR shows notable efficiency benefits due to its convergence-free nature \cite{henderson2000distributed}, it does not guaranty optimal path characteristics in terms of hop count or available bandwidth, resulting in comparatively lower throughput.
Therefore, our results demonstrate that the proposed MCTE method effectively captures throughput variations across different routing schemes, serving as a valuable benchmark to guide both routing scheme development and traffic optimization strategies.

\subsection{Impact of Traffic Pattern on Aggregate Throughput}

\begin{figure}[t!]
        \includegraphics[width=1\linewidth]{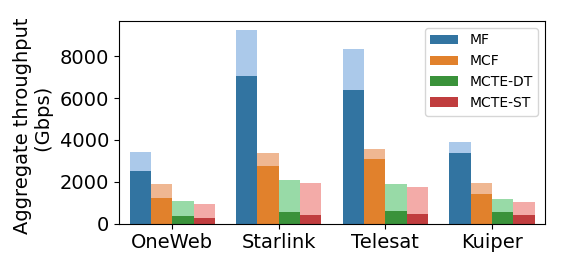}
    \caption{Throughput distribution of different traffic patterns under throughput estimation methods. The darker bars are under $\mathcal{D}_1$ traffic pattern while the lighter bars are under $\mathcal{D}_2$.} 
    \vspace{-2em}
        \label{fig:tfc_pattern}
 \end{figure}

Variations in traffic patterns can substantially impact network aggregate throughput performance. 
To quantify this effect, we analyze two distinct traffic matrices with identical total traffic volume but different spatial distributions:
\begin{itemize}
    \item  $\mathcal{D}_1$ employs a pure population model, where traffic generation probabilities between source-destination ground stations depend solely on city population density. The source-destination ground stations are paired randomly.
    \item  $\mathcal{D}_2$ incorporate both population and gravity models \cite{nucci2005problem}, introducing distance-dependent traffic generation where  shorter distance GSes pairs have higher occurrence probabilities
\end{itemize}
This formulation results in $\mathcal{D}_2$ traffic pattern exhibiting significantly shorter average communication distances compared to $\mathcal{D}_1$. \fig\ref{fig:tfc_pattern} demonstrates the performance variation of different throughput estimation methods under these traffic patterns across multiple constellations, with dark and light bars representing $\mathcal{D}_1$ and $\mathcal{D}_2$ respectively.
The results reveal that $\mathcal{D}_2$ consistently achieve superior throughput across all constellations and estimation methods, which illustrate that the reduction in inter-satellite switching can effectively decreases the resource consumption, thereby enabling LSN to support more communication sessions.

 \subsection{Impact of LSN Scale on Aggregate Throughput}

 We have previously evaluated only single-shell constellations. However, operational LSN systems now employ multiple orbital shells to improve both coverage and network capacity. To assess the impact of constellation scale, we use the MCTE method to evaluate Starlink’s throughput under increasing shell deployments.

\begin{figure}[t!] 
	\begin{center}

		\begin{minipage}{0.95\textwidth} 
		
		\begin{minipage}{1\textwidth} 
			\includegraphics[width=.52\textwidth]{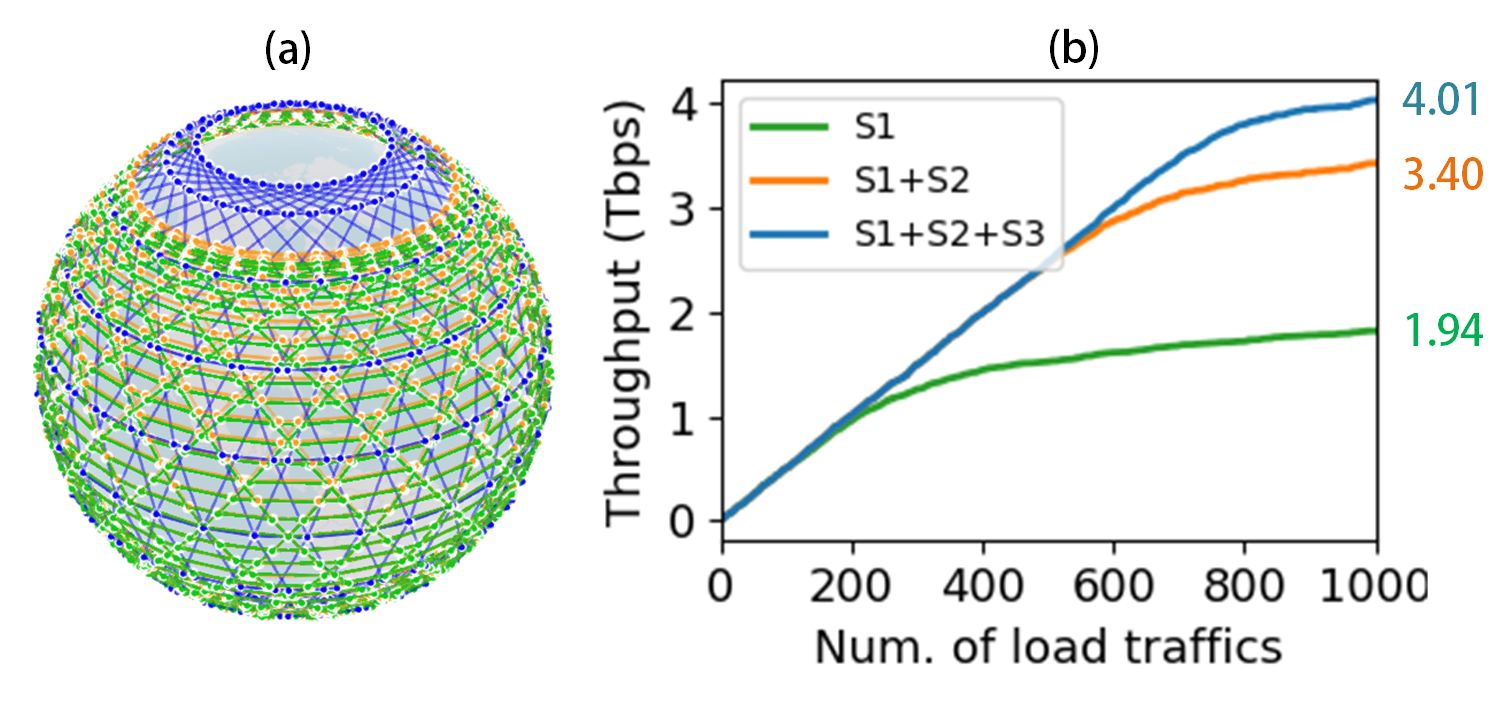} 
		  \end{minipage}  
	
		\end{minipage}  
		\scalebox{1}{
			\begin{tabular}{l|cccc}
			  \toprule[1pt]
				Starlink Gen1& Planes   & Sat/Plane    &Altitude&Inclination   \\  \toprule[1pt]
			  Shell-1& 22&72&550 km&$53^\circ$ \\
			  Shell-2& 22& 72&540 km&$53.2^\circ$ \\
			  Shell-3& 20   & 36 &570 km &$70^\circ$\\
			  \toprule[1pt]
		  \end{tabular}
			}
		
		\caption{Multi-shell Starlink constellation parameters \cite{starlink}, corresponding constellation scenario, and aggregate throughput curves under varying constellation scales.
		}
		\label{fig:scale} 
	\end{center}
\end{figure}

The table in \fig\ref{fig:scale} lists the parameters of Starlink’s three-shell constellation. In our evaluation, we incrementally add orbital shells to scale up the constellation size. Since no inter-shell ISLs are present, traffic flows are constrained to satellites within the same shell. All other traffic model parameters remain unchanged.
\fig\ref{fig:scale} (a) illustrates the constellation scenario, while \fig\ref{fig:scale} (b) presents the resulting throughput curves under increasing traffic load for each constellation scale.
We observe that, under identical traffic demand, throughput increases significantly as the network scales—an expected outcome of greater infrastructure capacity. Across all three scales, throughput grows linearly under low to moderate load, but eventually saturates as the network becomes bottlenecked. Notably, larger constellations sustain linear growth over a wider range of offered load and tolerate significantly more traffic before saturation.
At 1,000 traffic requests, the aggregate throughputs for the three scales are 1.94 Tbps, 3.40 Tbps, and 4.01 Tbps, respectively. This increase is not linear with the number of additional satellites, which we attribute to the diminishing marginal returns in terms of satisfied traffic requests as more links are occupied.

 \section{Conclusion}

This paper presents novel methods for capacity modeling and throughput estimation in dynamic LEO satellite networks. The proposed Cap-DLSN and MCTE approaches enhance both the realism and computational efficiency of throughput evaluation under unreliable ISL conditions. By capturing the inherent dynamics and topological variability of LEO constellations, these methods provide a practical and scalable framework for network performance assessment, serving as useful benchmarks for routing design, topology optimization, traffic engineering, and service-aware resource management in next-generation satellite networks. To support ongoing research, Cap-DLSN and MCTE will be open-sourced and integrated as built-in functions into the SNK \cite{snk}.



\section*{Conflict of interest}
The authors declare no potential conflict of interests.

\bibliographystyle{elsarticle-num}
\bibliography{ref}



\end{document}